\documentclass[chicago,numberedappendix,tighten]{emulateapj}

\usepackage{amsmath,amssymb,natbib,graphicx}
\usepackage{color}
\usepackage{ulem}

\slugcomment{DRAFT \today}

\shorttitle{Supernova Feedback Keeps Galaxies Simple}
\shortauthors{Chakraborti}

\begin{document}


\title{Supernova Feedback Keeps Galaxies Simple}


\author{Sayan Chakraborti}
\affil{Department of Astronomy and Astrophysics, Tata Institute of Fundamental Research,\\
    1 Homi Bhabha Road, Colaba, Mumbai 400 005, India}

\email{sayan@tifr.res.in}


%\altaffiltext{1}{Institute for Theory and Computation,
%Harvard-Smithsonian Center for Astrophysics, 60 Garden St.,
%Cambridge, MA 02138, USA}


\begin{abstract}
Galaxies evolve continuously under the influence of self-gravity, rotation,
accretion, mergers and feedback. The currently favored cold dark matter
cosmological framework, suggests a hierarchical process of galaxy formation,
wherein the present properties of galaxies are decided by their individual
histories of being assembled from smaller pieces. 
However, recent studies have uncovered surprising
correlations among the properties of galaxies, to the extent of forming a
one-parameter set lying on a single fundamental line.
It has been argued in the literature that such simplicity
is hard to explain within the paradigm of hierarchical galaxy
mergers. One of the puzzling results,
is the simple linear correlation between the neutral hydrogen mass and
the surface area, implying that widely different galaxies share very
similar neutral hydrogen surface densities. In this work we show that
self-regulated star formation, driven by the competition between
gravitational instabilities and mechanical feedback from supernovae,
can explain the nearly constant neutral hydrogen surface density
across galaxies. We therefore recover the simple scaling relation observed
between the neutral hydrogen mass and surface area. This result furthers
our understanding of the surprising simplicity in the observed properties
of diverse galaxies.
\end{abstract}





\section{Introduction}
In a hierarchical process of galaxy formation \citep{2000MNRAS.319..168C},
within the currently favored $\Lambda$-CDM cosmological
framework, the present properties of galaxies are decided by their individual
histories of being assembled from smaller halos \citep{1993MNRAS.262..627L}.
Moreover, these galaxies evolve continuously under the influence of self-gravity, rotation,
accretion and feedback.  
However, \citet{2008Natur.455.1082D} have recently reported surprising
correlations among the properties of galaxies, to the extent of forming a
one-parameter set lying on a single fundamental line \citep{2009MNRAS.394..340G}.
\citet{2008Natur.455.1049V} has argued that such simplicity
is hard to explain within the paradigm of hierarchical galaxy mergers.
One of the puzzling results
\citep{1983AJ.....88..881G,1984AJ.....89..758H,2001A&A...370..765V,2003ApJ...585..256R,2009MNRAS.394..340G},
is that the neutral hydrogen mass across these galaxies scales almost
linearly with surface area, implying that widely different galaxies
share very similar neutral hydrogen surface densities.
This nearly constant HI surface density has been pointed out in literature
to be an intriguing puzzle, demanding an explanation.

In this work we argue that correlation between the neutral hydrogen mass
and surface area may be preserved, despite the complex
merger histories of the galaxies, by self-regulated star formation,
driven by the competition between gravitational instabilities in the
rotating disk and mechanical feedback from supernovae. 
When mergers drive a galaxy away from the fundamental line, self
regulation of the porosity of the ISM and gravitational instability of
the star forming disk, as proposed by \citet{1997ApJ...481..703S},
can bring the neutral hydrogen surface density
back to the value which is predicted in our simple model.
This can explain
the regulation of the neutral hydrogen surface density in galaxies,
explaining part of the surprising simplicity in the observed properties
of galaxies.

\section{Surface density of gas in galaxies}
Evidence of a nearly universal neutral hydrogen (HI) surface density
in galaxies has been accumulating over the past three decades.
The initial hints came from single-dish 21 cm radio observations
of nearby galaxies. 
\citet{1983AJ.....88..881G} reported neutral hydrogen (HI) observations of 24 galaxies
in the Virgo cluster, using the Arecibo telescope.
The HI sizes and masses were found to be correlated in the same
manner, irrespective of whether they were HI rich or HI poor.
A similar correlation was soon found between the optical sizes and
HI masses of 288 isolated galaxies \citep{1984AJ.....89..758H}.
It was also found that the optical diameters of spiral disk are
better correlated with the HI mass than the morphological type
\citep{1984AJ.....89..758H}.
A deeper survey by \citet{2001A&A...370..765V} has revealed similar
correlations for galaxies in the Ursa Major Cluster.
The Arecibo Dual-Beam Survey \citep{2000ApJS..130..177R} (ADBS) has
found HI in 265 galaxies in a ``blind'' survey of $\sim430$deg$^2$ of sky.
While most of the ADBS galaxies were unresolved at the resolution
of the Arecibo, the Very Large Array (VLA) was used for interferometric
mapping of 84 galaxies and determine accurate sizes of 50 of them.
A comparison of HI masses and HI sizes of these along with 53 galaxies
with high resolution maps from literature revealed that they were 
consistent with a nearly constant average HI surface density of the order
$\sim10^7M_\odot$ kpc$^{-2}$ \citep{2003ApJ...585..256R}.
The HI masses of individual ADBS galaxies, spanning 3 orders of
magnitude, deviate by only $\sim0.13$ dex ($1\sigma$) from those
expected from a constant HI surface density. This puts the evidence, for a
regulated HI surface density across galaxies, on a firm observational basis.

A recent study of HI selected galaxies (free from optical selection effects)
found using the Parkes radio telescope and identified with SDSS sources,
has shown that 6 observed parameters, namely the dynamical mass
($M_d$), HI mass ($M_{HI}$), luminosity, color, and two radii
containing $50\%$ and $90\%$ of the observed luminosity, have 5
independent correlations among themselves \citep{2008Natur.455.1082D}.
This implies that the galaxies form a single parameter family and are
not removed significantly from their fundamental
line by the diverse merger histories that they would have had in
the process of hierarchical galaxy formation.
Some of the correlations have already been widely
known and discussed in other forms, such as the
correlation \citep{1996A&A...312..397G} between luminosity and dynamical mass.
\citet{2008Natur.455.1082D,2009MNRAS.394..340G} report a tight linear
correlation in the already established relation between the HI mass and
the surface area. The nearly constant HI surface density is argued in
literature to be an intriguing puzzle which demands an
explanation \citep{2009MNRAS.394..340G}. We discuss below, a physically
motivated explanation for this important observation.

\section{Gravitational instability in disk galaxies}
%The total SFR is determined \cite{1997ApJ...481..703S} by the interplay of P and
%the Toomre's \cite{1964ApJ...139.1217T} parameter Q.
Below a surface density threshold, azimuthally integrated star
formation in giant HII regions across a galaxy
ceases \citep{1989ApJ...344..685K}.
The existence of this threshold is traditionally stated in the
form of the Toomre parameter for gravitational instability
\citep{1968dms..book.....S}. Below a Toomre parameter $Q\lesssim1$,
rotation support and gas pressure cannot stabilize a thin self-gravitating
disk against gravitational instability
\citet{1960AnAp...23..979S,1964ApJ...139.1217T}.
Here $Q$ is defined as
\begin{equation}
 Q\equiv \mu _{cr} / \mu _{gas} ,
\end{equation}
where the critical gas surface density is given by
$\mu _{cr} \equiv \Omega \sigma_g / \pi G$, in terms of the angular
velocity $\Omega$ and velocity dispersion $\sigma_g$ in the gas disk.
\citet{2004ApJ...609..667S} has demonstrated that results from detailed
considerations including not only self-gravity, but metals, dust and
UV radiation, coincide with this empirically derived surface density
threshold for star formation.

%{\color{red}
This condition depends on the local angular velocity, which
we may define the  as $\mathbf{\Omega}=\nabla \times \mathbf{V}$.
Using cylindrical polar coordinates we can now write down the curl operator as
\begin{align}\nonumber
 \nabla \times \mathbf{A} = & \left({1 \over \rho}{\partial A_z \over \partial \phi}
 - {\partial A_\phi \over \partial z}\right) \boldsymbol{\hat \rho} +
 \left({\partial A_\rho \over \partial z} - {\partial A_z \over \partial \rho}\right)
 \boldsymbol{\hat \phi} \\ 
 & + {1 \over \rho}\left({\partial \left( \rho A_\phi \right) \over \partial \rho}   
 - {\partial A_\rho \over \partial \phi}\right) \boldsymbol{\hat z} .
\end{align}
Assuming, rotation and translation symmetry along $\boldsymbol{\hat \phi}$
and $\boldsymbol{\hat z}$ respectively, if the bulk motion of the gas is only
in the $\boldsymbol{\hat \phi}$ direction (as would be the case if the motion
is purely Keplarian) we may write down the angular velocity as
\begin{equation}
\Omega=\left|\mathbf{\Omega}\right| = \left|\nabla \times \mathbf{V} \right|
      = \left|{1 \over r}
      \left({\partial \left( r V (r) \right) \over \partial r}\right)
      \boldsymbol{\hat z}\right| ,
\end{equation} 
where the radial coordinate $\rho$ has been replaced by $r$, to avoid
confusion with density. In nearly rigid rotors such as dwarf galaxies,
the velocity is given by $V (r)=\Omega_0 r$. In such a case,
$\left|\nabla \times \mathbf{V} \right|$ is given by $2\Omega_0$
\citep[problem 2.4.7]{2005mmp..book.....A}. For flat ($V(r)=V_0$) rotation curves
 $\left|\nabla \times \mathbf{V} \right|=V_0/r$. Hence, for
all these cases, at boundary of the starforming disk, the local value
of the angular velocity is comparable to the global value. Hence, following
\citet{1997ApJ...481..703S} we shall use the global angular velocity
of the disk in the rest of this work.
%}

The
%{\color{red}
global
%}
angular velocity can now be expressed as
\begin{equation}
 \Omega_0=\sqrt{\frac{G M_d}{R^3}}
\end{equation}
in terms of the enclosed dynamical mass within the gas disk. Replacing
$M_d=4 \pi \rho_d R^3 / 3$, where $\rho_d$ is the nearly universal
dynamical mass density \citep{2008Natur.455.1082D,2009MNRAS.394..340G},
we have $\Omega_0=\sqrt{4 \pi G \rho_d / 3}$. Substituting, we get
\begin{equation}\label{mucr}
 \mu _{cr} = \frac{\sigma_g}{\pi} \sqrt{\frac{4 \pi \rho_d}{3 G}}.
\end{equation}
Now, we note that for the star forming disk to be long lived, it
may only be marginally unstable. Q is seen to be $\sim1$ throughout
the star forming regions in well observed disk
galaxies \citep{1989ApJ...344..685K}. This implies that throughout
the star forming disk the mean gas surface density is of order
$\mu _{gas} \sim \mu _{cr}$. Hence, the mean surface density of gas
is controlled primarily by the dynamical mass density $\rho_d$ and
the gas velocity dispersion $\sigma_g$.

One of the tight correlations seen by
\citet{2008Natur.455.1082D,2009MNRAS.394..340G} is
between the dynamical mass and cube of the optical radius, implying
a roughly constant average dynamical mass density (within the radii
of their star forming disks) for galaxies.
It is expected that Dark Matter contributes most of the gravitational mass
of the galaxies, hence a physical explanation of this observation would require
an understanding of the nature of Dark Matter.
Whether or not hierarchical halo build-up within a $\Lambda$-CDM cosmology
can explain this observation is still under investigation.
\citet{2010arXiv1011.6374L} have suggested that cold dark matter particles
interacting through a Yukawa potential could make halos beyond a certain
critical density evaporate over an Hubble time. This may set a characteristic
scale to the peak density of dark matter halos. Note however that the dark
matter halos of galaxies are much larger than the sizes of their star forming
disks. As a result, most of the dark matter may lie further out.
\citet{1987AJ.....93..816K} points out that the relative contributions of
the dark matter halo and stellar contents to the dynamical mass of a galaxy
vary significantly with its luminosity and morphological type. In the
rest of this work, we shall use the simple $M_d \propto R^3$ relation observed
by \citet{2008Natur.455.1082D,2009MNRAS.394..340G}, showing a
shared $\rho_d\sim10^7 M_\odot kpc^{-3}$ across galaxies.
This implies, that the mean surface density is controlled essentially
by the gas velocity dispersion $\sigma_g$ which is driven by energy
input from supernova explosions in the disk.

\section{Mechanical feedback from supernovae}
The distribution of gas in our galaxy has been variously described in
the past as being similar to that of Swiss Cheese \citep{1974ApJ...189L.105C},
as a Cosmic Bubble Bath \citep{1975A&A....38..363B} or as the Violent
Interstellar Medium \citep{1979ARA&A..17..213M}.
The basic idea is that supernova remnants
are full of coronal gas which can persist for long enough, as it radiates
very inefficiently by bremsstrahlung, so that a modest supernova rate may
produce an interconnected morphology of hot gas. Shells and supershells
\citep{1979ApJ...229..533H} have been found in the neutral hydrogen
distribution of galaxies. \citet{1987ApJ...317..190M} suggested that
stellar winds and repeated supernovae from an OB association may create
cavities of coronal gas in the interstellar medium leading to the formation
of supershells. \citet{2011ApJ...728...24C} have demonstrated
that this mechanism can explain the dynamics of a HI supershell found in
M101 driven by mechanical energy input from supernovae in a giant young
stellar association.

The fraction
of volume filled by supernova-driven hot gas is referred to as the porosity (P)
of the interstellar matter (ISM). The porosity is driven by the supply of hot gas
from supernova remnants (SNRs) and is related to the 4-volume of a SNR in the
cooling phase \citep{1988ApJ...334..252C} ($\nu_{SN}$) and the supernova rate
per unit volume ($r_{SN}$) as
\begin{equation}
 P=\nu_{SN} \times r_{SN}.
\end{equation}
The supernova rate is related to the recent star formation rate per 
unit volume ($\rho_{\ast}$) as
\begin{equation}
 r_{SN}=\rho_{\ast}/m_{SN},
\end{equation}
where $m_{SN}$ is the total mass of star formation required on an average 
to produce each core collapse supernova.
\citet{1997ApJ...481..703S} has argued that if P is too high,
it would suppress the efficiency of star formation. 
Since the supernova rate follows the recent star formation rate,
the situation called ``blowout'' would throttle the supply of hot gas and bring
down the value of P. On the other hand if P was too low, it would allow
the cold gas phase to dominate and form new stars more efficiently, some
of which would soon explode as supernovae and drive up the value of P.
Hence, there would be a self regulation process that would control P.

In this manner, \citet{1997ApJ...481..703S} shows that supernova explosions
supply the energy input necessary to maintain the velocity dispersion in
the gas phase at
\begin{equation}
 \sigma_g=6.90 P^{-0.58}n_g^{0.1}\, E_{51}^{0.2}\zeta^{0.008}\rm km\, s^{-1}.
\end{equation}
This expression depends only weakly on the gas density $n_g$, energy input
from individual supernovae $E_{51}\times10^{51}$ ergs, and the metallicity
$\zeta$ all of which have typical values of order unity.
The dependence on these parameters is dropped from our subsequent
equations. Assuming self-regulated star formation ($P\sim0.5$) the predicted
gas velocity dispersion has been shown \citep{1997ApJ...481..703S} to be
$\sigma_g\sim11$ km s$^{-1}$, close to the value of observed by
\citet{1989ApJ...339..763S} for the three-dimensional peculiar velocity
dispersion of interstellar molecular clouds within 3 kpc of the Sun.

\citet{2009ApJ...704..137J} have studied the effect of the supernova rate
on the interstellar turbulent pressure and confirmed a very weak dependence
of $\sigma_g$ on the star formation rate. They find simulated HI emission
lines widths of $10-18$ km s$^{-1}$ for models with SN rates that range from
1 to 512 times the Galactic SN rate.
Hence, the characteristic values for $\rho_d$ and $\sigma_g$ when plugged
into Equation \ref{mucr}, set the characteristic scale for the gas
surface density to a ballpark figure of $\sim10^7M_\odot$ kpc$^{-2}$, which
is close the observed value \citep{2003ApJ...585..256R} for the ADBS galaxies.

\begin{figure}
 \includegraphics[width=0.95\columnwidth]{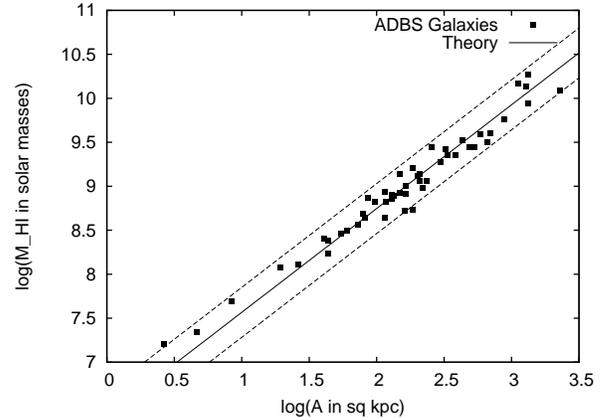}
 \caption{\textbf{Surface area and HI masses:} ADBS galaxies
\citep{2003ApJ...585..256R} are plotted as black squares. The solid
line is the theoretical curve (Equation \ref{M_HI}) from this work
and the area between the dashed lines represents the $2\sigma$ variance
region. Note the simple relation in which the HI mass scales almost
linearly with the surface area, implying a regulated HI surface density
across galaxies. This hitherto unexplained feature is reproduced by our
simple model.}
\end{figure}

\section{Regulated surface gas density}
Integrating the star formation rate per unit volume over an entire galaxy
\citet{1997ApJ...481..703S} provides the total SFR as 
\begin{equation}
\dot M_{\ast} \backsimeq 2.6 \times v_{rot,200}^{5/2} \, Q^{-3/2} \left( \frac{P}{0.5} \right)^{-0.87}\,\rm M_\odot\, yr^{-1},
\end{equation}
where $v_{rot,200}$ is the maximum rotational velocity of the galaxy,
normalized to 200 km s$^{-1}$. The exact expression \citep{1997ApJ...481..703S}
depends on well understood quantities like the total mass of stars formed to
yield one core collapse supernova, the mechanical energy output of an average
supernova, and depends only weakly on the metallicity.
As the disk becomes gravitationally unstable if $Q$ decreases, the SFR
increases and the resulting supernovae increase the porosity $P$. This
reduces the supply of cold gas and hence suppresses the star formation
rate. Hence \citet{1997ApJ...481..703S} points out that this expression
provides an explicit demonstration of disk self-regulation and
self-regulated star formation ensures $P\sim0.5$ and $Q\sim1$.
We use these values in the rest of this work.

Using the definition of the dynamical
mass ($M_d=(\Delta V)^2R_{90\%}/G)$) from \citet{1988gera.book...95K},
we can express $v_{rot,200}$ as
\begin{equation}
 v_{rot,200} \backsimeq 1.04 \times \sqrt{\frac{M_{d,11}}{R_{10}}},
\end{equation}
where $M_{d,11}$ is the dynamical mass normalized to $10^{11} M_\odot$ and
$R_{10}$ is the 90\% optical containment radius ($R_{90\%}$) normalized to 10 kpc.
The dynamical mass has been found to be correlated with the cube of the
optical radius \citep{2009MNRAS.394..340G}. The implied roughly constant
global dynamical density is of the order of $\rho_d\sim10^7 M_\odot kpc^{-3}$.
%Such a characteristic density scale may set the energy scale for interacting
%dark matter \cite{2010arXiv1011.6374L}.
Exploiting this observed relation we have
\begin{equation}
 M_{d,11}\backsimeq0.42 \rho_7 R_{10}^3,
\end{equation}
where $\rho_7=\rho_d / 10^7 M_\odot kpc^{-3}$.
Substituting for $v_{rot,200}$ and then for $M_{d,11}$, we get
\begin{equation}
\dot M_{\ast} \backsimeq0.95 \times \rho_7^\frac{5}{4} R_{10}^\frac{5}{2} \,\rm M_\odot\, yr^{-1}.
\end{equation}
%{\color{red}
One may interpret the \citet{1996A&A...312..397G} relation between total
luminosity (a proxy for the stellar mass) and dynamical mass (proportional
to $R^3$ because of the shared $\rho_d$) as $M_{\ast}\propto R^3$.
This implies that the doubling time for the stellar mass, scales as
\begin{equation}
 \tau=\frac{M_{\ast}}{\dot M_{\ast}} \propto R^{\frac{1}{2}}.
\end{equation}
Objects with larger doubling times have older stellar populations on
an average. This may explain why bigger galaxies are systematically redder.
%}

This
%{\color{red}
star formation rate
%}
allows us to estimate the mean SFR per unit area as
\begin{equation}
\Sigma_{SFR} \backsimeq3.0 \times 10^{-3} \times \rho_7^{1.25} R_{10}^{0.5} {\rm~M_\odot~yr^{-1}~kpc^{-2}}.
\end{equation}

However $\Sigma_{SFR}$ is related to the surface density of gas $\Sigma_{gas}$
by the Kennicutt-Schmidt Law \cite{1998ARA&A..36..189K} as
\begin{align}
\Sigma_{SFR} = & {{(2.5 \pm 0.7)} \times 10^{-4}}~
\\ \nonumber &
\times{\Big({\Sigma_{gas} \over 
   {1~M_\odot~{\rm pc}^{-2}}}\Big)^{1.4\pm0.15}}{\rm~M_\odot~yr^{-1}~kpc^{-2}}.
\end{align}
If the molecular gas fraction is $\ll1$, which is true for most disk galaxies,
the mean surface density of HI $\Sigma_{HI}$
is comparable to the total gas surface density $\Sigma_{gas}$.
This assumption will make a proportional error comparable to the
molecular gas fraction.
Hence assuming $\Sigma_{HI}\sim\Sigma_{gas}$,
eliminating the $\Sigma_{SFR}$ and integrating over the surface area,
we obtain the relation between the gas mass and the surface area as
%\begin{equation}
% log\Big(\frac{M_{HI}}{1 M_\odot}\Big) = (6.91 \pm 0.08)
% + (1.18\pm0.04) log\Big(\frac{A}{1 kpc^2}\Big)
% + (0.89\pm0.10) log\Big(\frac{\rho_d}{10^7 M_\odot kpc^{-3}}\Big).
%\end{equation}
\begin{align}
\nonumber
 log\Big(\frac{M_{HI}}{1 M_\odot}\Big) \backsimeq & (6.91 \pm 0.08)
 + 1.18 \times log\Big(\frac{A}{1 kpc^2}\Big)
\\ &
+ 0.89 \times log\Big(\frac{\rho_d}{10^7 M_\odot kpc^{-3}}\Big),
\label{M_HI}
\end{align}
where $A\equiv\pi R^2$ is the cross sectional surface area presented
by the galaxy.
Given that dynamical mass density ($\rho_d$) is not seen to vary much across
galaxies, the almost linear relation between the total HI mass and the surface
area implies very similar HI surface densities across a range of
galaxies. The normalization matches the
observed \citep{2003ApJ...585..256R} relation for $\rho_7=0.26$ which lies within
the range of its observed \citep{2009MNRAS.394..340G} values. The HI masses,
of ADBS galaxies \citep{2003ApJ...585..256R}, spanning 3 orders of magnitude,
vary only by $\sim0.14$ dex ($1\sigma$) from the theoretical curve. However,
this scatter is larger than the scatter propagated from the Kennicutt-Schmidt
Law \citep{1998ARA&A..36..189K}. The excess scatter may be attributed to the
scatter in core properties of halos such as $\rho_7$. It has been pointed out
by \citet{2010arXiv1011.6374L} that numerical simulations are
required to determine the scatter in dark matter core densities as a function
of mass and redshift. This would be important for dark matter dominated halos.
\citet{1996A&A...312..397G} show a correlation between total luminosity and
the dynamical mass. If most of the dynamical mass is provided by the stellar
content, it would be important to study the scatter in this relationship.

\section{Discussions}
Our result shows that the present understanding of mechanical
feedback from supernovae, leading to self regulated star formation
\citep{1997ApJ...481..703S}, can account for the regulation of HI
surface density across galaxies to a characteristic value of around
$\sim10^7M_\odot$ kpc$^{-2}$ \citep{2003ApJ...585..256R}.
The predicted HI surface density depends on the mean density of the
dynamical mass, which is likely to be provided by a combination of
dark matter and stellar content.
As to why galaxies are observed \citep{2009MNRAS.394..340G} to share
similar dynamical mass densities is still an open case requiring further
investigation. Cold dark matter particles interacting through a Yukawa
potential \citep{2010arXiv1011.6374L} could provide a natural explanation
for a characteristic density in dark matter dominated halos.
The correlation \citep{1996A&A...312..397G} between luminosity
and dynamical mass may be important in halos dominated by the stellar content.

%{\color{red}
The model of self regulated star formation \citep{1997ApJ...481..703S}
has been shown in this work to provide a scaling between doubling time
and radius. This could explain why larger galaxies are systematically redder.
This relation should be used in conjunction with population synthesis models
such as Starburst99 \citep{1999ApJS..123....3L}
to predict colors as a function of galaxy size. This could provide an
interesting test of this model in future.
%}
 
Even if mergers drive a galaxy away from the fundamental line, self
regulation of the porosity of the ISM and Toomre parameter of the star forming
disk can bring the HI surface density back to the value which is
predicted in our simple model and observed in a wide range of galaxies.
In a framework for hierarchical galaxy
mergers this can happen multiple times in a galaxy's history, until it
eventually runs out of neutral hydrogen. A detailed understanding of
the proposed mechanism would require self consistent galaxy simulations
taking into account the supply of hot gas into the ISM from
individual supernovae.

Our model relates the total neutral hydrogen mass of the galaxy with
its projected surface area. However in practice, the quantities observed
with radio telescopes are the redshift, integrated HI line fluxes and
the solid angles on the sky.
Fluxes and solid angles behave differently in different cosmological
scenarios, as they scale with the luminosity distance $d_L$ and angular
diameter distance $d_A$ respectively. This could facilitate their use in
testing the \citet{1933PMag...15..761E} relation, $d_L = (1 + z)^2 d_A$,
when future telescopes such as the Square Kilometer
Array start to detect red-shifted HI from very distant galaxies.


%% Included in this acknowledgments section are examples of the
%% AASTeX hypertext markup commands. Use \url without the optional [HREF]
%% argument when you want to print the url directly in the text. Otherwise,
%% use either \url or \anchor, with the HREF as the first argument and the
%% text to be printed in the second.

\acknowledgments

The author wishes to thank thank Alak Ray for guidance, a careful
reading of the manuscript and numerous suggestions. Abraham Loeb,
Swastik Bhattacharya and Satej Khedekar are thanked for discussions
about the work. An anonymous referee is thanked for suggestions.

% PUT FIGURES HERE




%\clearpage

\bibliographystyle{apj}
\bibliography{simple_galaxies}

%\clearpage

%% the file, uncomment the \plotone, \plottwo, and
%% If you need a layout that cannot be achieved with \plotone or
%% \plottwo, you can invoke the graphicx package directly with the
%% \includegraphics command or use \plotfiddle. For more information,
%% please see the tutorial on "Using Electronic Art with AASTeX" in the
%% documentation section at the AASTeX Web site,
%% http://www.journals.uchicago.edu/AAS/AASTeX.
%%
%% The examples below also include sample markup for submission of
%% supplemental electronic materials. As always, be sure to check
%% the instructions to authors for the journal you are submitting to
%% for specific submissions guidelines as they vary from

%% This example uses \plotone to include an EPS file scaled to
%% 80% of its natural size with \epsscale. Its caption
%% has been written to indicate that additional figure parts will be
%% available in the electronic journal.

\end{document}